\documentclass[aps,prd, superscriptaddress,preprintnumbers,nofootinbib]{revtex4}
\usepackage{amsmath,amsfonts,amssymb,amsthm,amstext,amscd,eucal,graphicx , xcolor}
\usepackage[all]{xy}
\usepackage[colorlinks=true]{hyperref}
\usepackage{color}
\usepackage{amsmath}
\usepackage{bigints}
\def\be{\begin{equation}}
\def\ee{\end{equation}}
\def\arr{\begin{array}{rll}}
\def\ea{\end{array}}
\def\bea{\begin{eqnarray}}
\def\eea{\end{eqnarray}}
\def\N2{$N{=}2$}

\def\>{\rangle}
\def\<{\langle}
\def\+{\dagger}
\def\={\ =\ }

\begin{document}

\preprint{IPM/P-2017/009}
\vskip 1 cm
\title{Near horizon extremal Myers-Perry black holes\\
and integrability of associated conformal mechanics}
\author{Tigran Hakobyan}
\email{tigran.hakobyan@ysu.am}
\affiliation{Yerevan State University, 1 Alex Manoogian St., Yerevan, 0025, Armenia}
\affiliation{Tomsk Polytechnic University, Lenin Ave. 30, 634050 Tomsk, Russia}
\author{Armen Nersessian}
\email{arnerses@ysu.am}
\affiliation{Yerevan State University, 1 Alex Manoogian St., Yerevan, 0025, Armenia}
\affiliation{Tomsk Polytechnic University, Lenin Ave. 30, 634050 Tomsk, Russia}
\author{M.M. Sheikh-Jabbari}
\email{jabbari@theory.ipm.ac.ir }
\affiliation{ Institute for Research in Fundamental Sciences (IPM), P.O.Box 19395-5531, Tehran, Iran   }
\begin{abstract} We investigate  dynamics of probe particles moving in the near-horizon limit of $(2N+1)$-dimensional extremal Myers-Perry black hole with arbitrary rotation parameters. We observe that in the most general case with nonequal nonvanishing  rotational parameters the system  admits separation of variables in $N$-dimensional ellipsoidal coordinates. This allows us to find  solution of the corresponding Hamilton-Jacobi equation and write down the explicit expressions of Liouville constants of motion.
\end{abstract}
\maketitle


%\renewcommand{\thefootnote}{\arabic{footnote}}
\setcounter{footnote}0
\section{Introduction and review}

Analyzing causal curves, timelike or null geodesics, is the pivotal part of any black hole probe study. These geodesics carry the information about black hole charges and their dynamics. In particular, for realistic Kerr-type black holes geodesics probing the region close to the event horizon of black hole is an essential part of studying black hole accretion disks or black hole mergers. Moreover, among the black holes in the sky there are very fast rotating black holes which can be well approximated by an extremal Kerr geometry \cite{Extreme-Kerr}. Focusing on the near horizon region, relevant to the accretion disk dynamics, it has been argued that for the extremal black holes one can analytically solve the associated plasma equations in the physically relevant limit of force-free electrodynamics where energy momentum of the electromagnetic field dominates over that of the charged matter fields \cite{force-free-1,force-free-2}.

Besides the direct observational motivations, extremal black holes and their geodesics have been of great interest for more general class of black holes. In this work we will be focusing on the geodesics probing the near horizon region of Extremal Myers-Perry (EMP) black holes \cite{mp}, which are higher dimensional counterparts of extremal Kerr black hole. In general  geodesic equation is basically describing a
$d$ dimensional particle dynamics in certain potential. The relevant question is then exploring integrability of this dynamical system.

Various aspects of extremal black holes, black hole which have vanishing surface gravity or have a degenerate (non-bifurcate) horizon, have been studied. These black holes usually have the lowest possible mass for a given set of angular momentum or other charges and have the remarkable property that at the near horizon there is an enhancement of isometries. There are theorems that for stationary extremal black holes the $U(1)$ isometry associated with the horizon generating Killing vector field in the near horizon (NH) region enhances to a three dimensional group (associated with three Killing vector fields) and that this NH isometry group is generically $SO(2,1)\simeq SL(2,\mathbb{R})$ \cite{bh,NHEG-general,NHEG-2,KL-review}. Since $SL(2,\mathbb{R})$ is the one dimensional conformal group, the particle dynamics on the near horizon extreme geometries possesses  dynamical conformal symmetry, i.e.  defines ``conformal mechanics". This brings the hope of making general statements on the integrability of the system of question, see e.g.
see \cite{conformal-mechanics-BH-1,conformal-mechanics-BH-2, Anton-MP, Anton-1, GNS-1} and references therein.

 The dynamical $SL(2,\mathbb{R})$ invariance allows  performing canonical transformation under which the  Hamiltonian of the system  formally takes the non-relativistic form \cite{conformal-mechanics-BH-2,GNS-1}
\be\label{nonrel}
H=\frac{p^2_R}{2}+\frac{2\mathcal{I}}{R^2},
\ee
where
\be\label{R-p_R}
R=\sqrt{2K}, \quad p_R=\frac{2D}{\sqrt{2K}}
\ee
are  the effective ``radius" and its canonical conjugate ``radial momentum", and $\mathcal{I}=HK-D^2$ is the Casimir of the $SL(2,\mathbb{R})$  algebra whose generators $H,D,K$ satisfy the relataions
\be
\{H,D\}=H, \quad \{H,K \}=2D, \quad \{D,K \} =K.
 \label{confalg}\ee
The $SL(2,\mathbb{R})$ Casimir  $\mathcal{I}$ depends on the $d-1$  ``angle-like" variables and their conjugate momenta which  commute with ``radial variables" $R,p_R$.  All specific
properties of such systems are hence encoded in $\mathcal{I} $ which in turn may be viewed as the Hamiltonian of another associated system. Such associated systems have been investigated from  various viewpoints  where they were called ``angular (or spherical) mechanics" , see  \cite{Armen-Tigran} and refs therein.

Although the spherical mechanics related to nonrelativistic conformal  models has been extensively studied, systems originating from near horizon extremal black holes received less attention \cite{conformal-mechanics-BH-2,Anton-MP,GNS-1,Anton-1}.
In particular, a special class of NH geometry of $d$ dimensional EMP black holes with  $SO(2,1)\times U([\frac{d-1}{2}])$  isometry group
was considered in \cite{Anton-MP,GNS-1}. It was found  that in the odd dimensions, $d=2N+1$, the angular mechanics part reduces to the $(N-1)$- dimensional singular  spherical oscillator and established that it is   superintegrable system, i.e. possesses maximal number, $2N-3$ of constants of motion  \cite{GNS-1}.
For the even, $d=2N+2$ cases, it was shown that  the angular mechanics is an $N$- dimensional    integrable system with $2N-2$ constants of motion,
 containing latter one as a subsystem. Thus, it loses  maximal  superintegrability feature.

In this work we revisit the case of particle dynamics in the near horizon extremal Myers-Merry (NHEMP) black holes in $2N+1$ dimensions and consider the most general case where the isometry of the background is $SL(2,\mathbb{R})\times U(1)^N$ and explore the integrability of the system. As in the 5d NHEMP case, we do not expect the system to be superintegrable \cite{GNS-1}. The questions we will address in this work are
\begin{itemize}
\item Is its spherical mechanics part an integrable system, and if so does it admit separation of variables?
Given that the geodesics of  general  higher-dimensional black hole metrics admits separation of variables  \cite{frolov}, we expect the answer to this question to be positive.
\item Are there special values of rotational parameters when the system gets additional constant(s) of motion?
\end{itemize}

The main results of our study are:
 \begin{itemize}
 \item We establish that the angular mechanics  $\mathcal{I}$ in general admit  separation of variables in $N$-dimensional ellipsoidal coordinates and find the explicit expressions of its generating function and
  the Liouville constants of motion.
%\item {\color{red} Hamilton-Jacobi equation (and  Schr\"odinger equation in quantum case) of the system written in ellipsoidal coordinates corresponds to that of
% $N$-dimensional oscillator. Therefore, the system   is not only integrable, but exactly solvable. {\tt now I am not sure, that it is true. Probably, we should remove this statement.}}
\item Having the example of equal angular momentum parameters \cite{GNS-1} in mind, one can argue that  there exists  specific values of rotational parameters the system possesses higher order constants of motion and  is superintegrable.
 \end{itemize}

The paper is organized as follows. In  {\sl Section 2}  we represent the $(2N+1)$ dimensional NHEMP geometry \cite{NHEG-2} in coordinates convenient for our study, and then
 reformulate the particle dynamics on it in the $N$-dimensional ellipsoidal coordinates. In  {\sl Section 3} we write down the conformal mechanics describing the motion of a probe particle in this background and construct the associated ``angular mechanics". We show that the corresponding Hamilton-Jacobi equation separates in the ellipsoidal coordinates and find its solution. Using this solution we construct the explicit expressions of the Liouville constants of motion. The last section is devoted to concluding remarks. Some of the technical details are gathered in two appendices.

\section{NHEMP metrics}
Myers-Perry black holes \cite{mp} are $d$ dimensional, asymptotic flat, Einstein vacuum solutions. For $d=2N+1$ case these solutions come with $N+1$ parameters, $N$ angular momentum/velocity parameters  and a mass parameter and in the extremal case the mass parameter is given in terms of the angular momentum parameters. The NHEMP metric is given in the appendix \ref{appendix} and in the appropriate parametrization takes the form
\be
\frac{ds^2}{r^2_H}=A(x)\left(-r^2d\tau^2+\frac{dr^2}{r^2}\right)+\sum_{i=1}^N dx_idx_i+
\sum_{i,j=1}^N\tilde{\gamma}_{ij}x_i x_j D\varphi^iD\varphi^j,\qquad D\varphi^i\equiv d\varphi^i+k^ird\tau
\label{m2}\ee
where $r_H$ is the horizon radius of the original black hole,
\begin{subequations}\begin{align}
A(x) =&\frac{\sum_{i=1}^N x^2_i/m^2_i}{1+4\sum_{i<j}(m_i m_j)^{-1}},\label{A}\\
\tilde{\gamma}_{ij}=&\delta_{ij}+ \frac{1}{\sum_l x_l^2/m^2_l}\frac{\sqrt{m_i-1}x_i}{m_i}  \frac{\sqrt{m_j-1}x_j}{m_j},
\label{gamma}\\
k^i&=2\frac{{\sqrt{m_i-1}/m^{2}_i}}{1+4\sum_{l<n}(m_lm_n)^{-1}},\label{k}
\end{align}\end{subequations}
while $m_i\geq 1$ and $0<x_i\leq \sqrt{m_i} $,  and  obey the conditions\footnote{Note that NHEMP is an Einstein vacuum solution and hence $r_H$ is not determined by the Einstein equations of motion. This solution, besides $r_H$, has  hence $N-1$ independent parameters.}
\be
\sum_{i=1}^N \frac{x^2_i}{m_i}=1,\quad \sum_{i=1}^N\frac 1m_{i}=1.\label{ell}
\ee
That is, $x_i$ can be interpreted as an ambient Cartesian coordinates of the $(N-1)$-dimensional ellipsoid with $\sqrt{m_i}$ semiaxes.  In this paper we focus on generic EMP case where neither of $m_i$ are  equal to one. Without loss of generality we  can choose $1<m_N\leq m_{N-1}\leq m_{N-2}\leq\ldots \leq m_1$, leading to $m_N \leq N\leq m_1$.

The relation of these coordinates and parameters with conventional latitudinal coordinates and rotational parameters of black hole are presented in the Appendix \ref{appendix}. When the rotational parameters coincide, $m_i={N}$, the Hamiltonian of probe particle  reduces to the system on sphere and admits separation of variables in spherical coordinates \cite{GNS-1}.
Noting the metric \eqref{m2} and \eqref{ell},
it seems plausible that in the $N$-dimensional ellipsoidal coordinates
the respective dynamics  admits separation of variables. We will show below that this is indeed the case.

To this end, let us first assume that $m_i$'s generic, neither of them are equal, and introduce the coordinates $\lambda_i$
\be
x^2_i=(m_i-\lambda_i)\prod_{j=1, j\neq i}^{N}\frac{m_i-\lambda_j}{m_i-m_j},
\label{xN}\ee
where $\lambda_N  < m_N  <  \ldots < \lambda_2  < m_2  < \lambda_1 < m_1$.
In these  coordinates
\be\label{m0}
\sum_{i=1}^N dx_i^2=\sum_{i=1}^N h_i^2(\lambda)d\lambda_{i}^{2},
\qquad{\rm where}
\qquad
h_i^2=\frac{\prod_{j=1,j\neq i}^{N}(\lambda_j-\lambda_i)}{4\prod_{j=1}^{N} (m_j-\lambda_i)}.
\ee

The key to separation of variables is the interesting identity,
\be
\sum_{i=1}^N \frac{x^2_i}{m^2_i}=\left(\prod_{i=1}^N{\frac{\lambda_i}{m_i}}\right)\left(\sum_{i=1}^N\frac{1}{\lambda_i}-1\right),
%,\qquad\sum_{i=1}^N \frac{\lambda^{N-1}_i}{\prod_{i\neq j}^N(\lambda_i-\lambda_j)}=1.
\label{12}\ee
and  that the constraint \eqref{ell} can be solved through $\lambda_N=0$. To work out the above we have used identities in the Appendix \ref{appendix-useful}.
%\footnote{The constraint \eqref{ell} leads to $\lambda_1\lambda_2\cdots \lambda_N=0$.}
This restricts
$N$-dimensional Euclidean metrics  \eqref{m0} to the metrics on $(N-1)$-dimensional ellipsoid\footnote{As is implicit, in our notation $a,b$ indices run over $1,\cdots, N-1$ and $i,j$ over $1,\cdots, N$.}
\be
 h_{ab}d\lambda_a d\lambda_b=-
\sum_{a=1}^{N-1}
 \frac{ \prod_{b=1,a\ne b}^{N-1}(\lambda_b-\lambda_a)\lambda_a d\lambda_a^2}{4\prod_{i=1}^{N}(m_i-\lambda_a) },
\label{hab}
\ee
and hence
\be\label{A-lambda}
A=\frac{m_N^{-1}}{1+4\sum_{i<j}(m_i m_j)^{-1}}\left(\prod_{a=1}^{N-1}{\frac{\lambda_a}{m_a}}\right).
\ee
With these expressions at hand we are ready to consider the particle dynamics in the given background \eqref{m2}.

\section{Conformal mechanics}\label{sec-conf-mech}
In the above notation the mass-shell equation for a particle of mass $m_0$ moving in the background metrics \eqref{m2} reads
\be\label{kg}
m^2_0=\sum_{A,B=1}^{2N+1}g^{AB}p_A p_B
 \quad
 \Leftrightarrow\
\quad
 m_0^2r^2_H+\sum_{a,b,d=1}^{N-1}h^{ab}{\pi_a}{\pi_b}+\sum_{i,j=2}^N{\widetilde\gamma}^{ij}\frac{p_i}{x_i}\frac{p_j}{x_j}=\frac{1}{A}\left[ \left( \frac{ p_0}{r}- \sum_i{k}_i p_i\right)^2 -({rp_r})^2 \right]
\ee
where $h^{ab}$ is the inverse metrics to \eqref{hab},
\be
{\tilde\gamma}^{ij}=\delta^{ij} - \frac{x_i\sqrt{m_i-1}}{m_i} \frac{x_j\sqrt{m_j-1}}{m_j} ,
\ee
and $\pi_a$ are conjugate momenta to $\lambda_a$ with the canonical Poisson brackets
\be\label{Poisson-bracket}
\{\pi_a, \lambda_b\}= \delta_{ab },\qquad \{p_i,\varphi^j\}=\delta_{ij},\qquad\{p_r, r\}=1.
\ee
From the  expression \eqref{kg} we get the explicit form of Hamiltonian
\be
H=p_0= r\left(\sqrt{L(\pi_a,x_a, p_i)+(rp_r)^2}+ \sum_i k_ip_i\right),
\label{ho}\ee
where
\be
L=A\left[ \sum_{a,b,d=1}^{N-1}h^{ab}{\pi_a}{\pi_b}+\sum_{i,j=2}^N{\widetilde\gamma}^{ij}\frac{p_i}{x_i}\frac{p_j}{x_j} +m_0^2r^2_H \right],
\ee
and consequently, recalling analysis of \cite{conformal-mechanics-BH-2,Anton-MP,Anton-1}, we obtain the expressions  for the generators of conformal boost $K$ and of the dilatation $D$ which obey the algebra \eqref{confalg},
\be\label{triple}
 D=r p_r,\qquad
K=\frac{1}{r} \left( \sqrt{{(r p_r)}^2 + L(x_a,\pi_a,p_i)}- \sum_i k_ip_i
 \right).
%\ee\be
%\label{confalg}
%\{H,D\}=H, \quad \{H,K \}=2D, \quad \{D,K \} =K,
\ee
Hence, the Hamiltonian \eqref{ho} can be represented in formally nonrelativistic form \eqref{nonrel}.\footnote{
Note that the radial variables $(R, p_R)$ \eqref{R-p_R} do not commute (with respect to the Poisson brackets \eqref{Poisson-bracket}) with $p_\mu=(\pi_a, p_i), q_\mu=(\lambda_a,\varphi^i)$.
In order to split them, one  can perform a canonical transformation
$(r,p_r, p_\mu, q^\mu)\to
(R,p_R, {\widetilde q^\mu}, {\widetilde p_{\mu}})$, which is defined by \eqref{R-p_R}
and by an
appropriate  transformation of the remaining variables \cite{conformal-mechanics-BH-2,GNS-1}.}

%\be \label{confalg}
%\{H,D\}=H, \quad \{H,K\}=2D, \quad \{D,K\}=K.
%\ee
The Casimir element of conformal algebra then reads
\begin{gather}
\mathcal{I}=H K-D^2=L-\left (\sum_i k_ip_i\right)^2
=A\left[      \sum_a h^{ab}\pi_a\pi_b
+\sum_i\frac{p^2_i}{x^2_i} +g_0
\right]- \mathcal{I}_0  ,
\label{L}\end{gather}
where $A$ is given in \eqref{A-lambda}, $x^2_i$ are given by \eqref{xN} with $\lambda_N=0$, and
\be\label{consts}
%\begin{gathered}
%C_{N}=\frac{1}{\left(1-8\sum_{i<j}(m_i m_j)^{-1}\right)m_1\ldots m_N},
%\qquad
 g_0={ -\left(\sum_{i=1}^N \frac{\sqrt{m_i-1}p_i}{m_i}\right)^2}+m^2_0r^2_H,
 \qquad
 \mathcal{I}_0=4\left(\sum k^ip_i\right)^2\ .
 % \left(\sum_i \frac{2}{1-8\sum_{i<j}(m_i m_j)^{-1}}\frac{\sqrt{m_i-1}p_i}{m^2_i}\right)^2}.
%\end{gathered}
\ee
The above provides an explicit representation of our system in the ``non-relativistic form'' \eqref{nonrel}.
As we see the Casimir   $\mathcal{I}$ \eqref{L} is at most quadratic in momenta canonically conjugate to the remaining angular variables and it can conveniently be viewed as the Hamiltonian of a reduced ``angular/spherical mechanics"  describing motion of particle on some curved background.
Since  the azimuthal angular variables $\varphi^i$ are cyclic, corresponding conjugate momenta $p_i$ are constants of motion. We then remain with a reduced $N-1$
dimensional system described by Hamiltonian \eqref{L} and $\lambda_a$ variables and their conjugate momenta. The reduced Hamiltonian with \eqref{consts} and $p_i$ as (coupling) constants is
\be
\tilde{\mathcal{I}}=\lambda_1\ldots\lambda_{N-1}\left[ - \sum_a\frac{{4\prod_{i=1}^{N}(m_i-\lambda_a) }\pi^2_a}{\lambda_a\prod_{b=1,a\ne b}^{N-1}(\lambda_b-\lambda_a)}
+\sum_{i=1}^N\frac{{g}^2_i}{\prod_{a=1}^{N-1}(m_i-\lambda_a)} +g_0\right],
\label{24}\ee
where we introduce further notation
\be{ g}^2_i =\frac{p^2_i}{m_i}\prod_{j=1,j\neq i}^{N} (m_i-m_j),\qquad \tilde{\mathcal{I}}\equiv \frac{\mathcal{I}+\mathcal{I}_0}{C_{ N}}, \qquad 
C_N=\frac{A}{\lambda_1\ldots\lambda_{N-1}}={\frac{1}{\left(1+4\sum_{i<j}(m_i m_j)^{-1}\right)m_1\ldots m_N}}.
\ee
Using the identities in the Appendix \ref{appendix-useful}, we can rewrite the Hamiltonian
expression  \eqref{24} in an implicit form:
\be\label{29}
\sum_{a=1}^{N-1} \frac{R_a-\tilde{\mathcal{I}}\lambda^{-1}_a}{\prod_{b=1,a\ne b}^{N-1}(\lambda_b-\lambda_a)}=0,
\ee
where
\be
 R_a \equiv R(\lambda_a, \pi_a )=-4\prod_{i=1}^{N}{(m_i-\lambda_a)} \frac{\pi^2_a}{\lambda_a}
 +(-1)^N \sum_{i=1}^N\frac{{ g}^2_i}{m_i-\lambda_a}+g_0(-\lambda_a)^{N-2}.
\label{30}
\ee
Equipped with the above, we can solve the  Hamilton-Jacobi equation
$$\tilde{\mathcal{I}}(\lambda_a,\partial S_{gen}/\partial \lambda_a )=\mathcal{E}
$$
and obtain the generating function $S_{gen}$ depending on $N-1$ integration constants. To this end,
 noting \eqref{30}, one can show that
\be\label{S-gen}
S_\text{gen}(\lambda_1,\dots,\lambda_{N-1})=\sum_{a=1}^{N-1} S(\lambda_a).
\ee
Using the identity \eqref{28}, the solution of \eqref{29} which depends on $\mathcal{E}$ and
$N-2$ integration constants $\nu_{\alpha}$ is given through
\be
\label{31}
R\Big(\lambda_a,\frac{d S(\lambda_a)}{d\lambda_a}\Big)-\frac{\mathcal{E}}{\lambda_a}
-\sum_{\alpha =1}^{N-2}\nu_{\alpha}\lambda^{\alpha -1}_a=0,
\ee
or in an  explicit form,
\be
-{4}\left(\frac{dS(\lambda_a)}{d\lambda_a}\right)^2\prod_{i=1}^{N}{(m_i-\lambda_a)}
+ (-1)^N\sum_{i=1}^N\frac{{ g}^2_i {\lambda_a}}{m_i-
\lambda_a}+g_0(-\lambda_a)^{N-1}-{\mathcal{E}}-\sum_{\alpha =1}^{N-2}\nu_{\alpha}\lambda^{\alpha}_a=0.
\label{part}\ee
Hence,  the analytic solution to the Hamilton-Jacobi equation is given through the generating function \eqref{S-gen}  with
\be\label{Generating-function}
%\begin{aligned}
%S_\text{gen}&=\sum_{a=1}^{N-1} S(\lambda_a,\nu_\alpha,\mathcal{E}),
%\\
S(\lambda,\nu_\alpha,\mathcal{E})=\frac 12\bigintsss\frac{d\lambda}{\sqrt{
\prod_{i=1}^{N}{(m_i-\lambda)}}}
\sqrt{
(-1)^N\left[\sum_{i=1}^N\frac{{ g}^2_i m_i}{m_i-\lambda}
+g_0\lambda^{N-1}-\sum_{i=1}^N{ g}^2_i\right]
-\sum_{\alpha =1}^{N-2}\nu_{\alpha}\lambda^{\alpha} -{\mathcal{E}}}\ .
%\end{aligned}
\ee

From \eqref{31}  we can get the analytic expressions of the commuting  constants of motion. For this purpose, we  represent it  in a more compact form as
\be
\sum_{\alpha =0}^{N-2}\nu_{\alpha}\lambda^{\alpha}_a=\lambda_aR_a(\pi_a,\lambda),
\qquad
\text{where}
\qquad
\nu_0=\mathcal{E}.
\ee
which may be rewritten in terms of the Vandermonde matrix $W$,
\be
\begin{pmatrix}
1 & \lambda_1 & \lambda_1^2 & \cdots & \lambda_1^{N-2}
\\
1 & \lambda_2 & \lambda_2^2 & \cdots & \lambda_2^{N-2}
\\
\vdots & \vdots & \vdots & \ddots &\vdots
\\
 1 & \lambda_{N-1} & \lambda_{N-1}^2 & \cdots & \lambda_{N-1}^{N-2}
\end{pmatrix}
\begin{pmatrix}
\nu_0
\\
\nu_1
\\
\vdots
\\
\nu_{N-2}
\end{pmatrix}
=
\begin{pmatrix}
\lambda_1 R_1
\\
\lambda_2 R_2
\\
\vdots
\\
\lambda_{N-1} R_{N-1}.
\end{pmatrix}.
%\    & = \prod_{0\leq j < i \leq n} (\lambda_{i} - \lambda_{j})
\ee
The solution  may then be expressed via the inverse Vandermonde matrix $W^{-1}$,
which exists for distinct set of $\lambda_a$, ($\lambda_a\neq \lambda_b$ if $a\neq b$).
Then using  these equations, we can find the expressions of $\nu_\alpha$
via momenta and coordinates, which defines the Liouville constants of motion.
{\sl Hence, we proved the integrability of the integrability of the system under consideration.}
Notice, that the partial Hamilton-Jacobi equation \eqref{part} corresponds  to  those of $N$-dimensional oscillator.
The same is true  for its quantum counterpart, Schr\"odinger equation. Hence, one can expect that the system of question is not only integrable, but also exactly solvable.

Let us conclude this section by the few words about hidden symmetries and superintegrability. While the generic system is clearly not superintegrable,   for the specific values of rotational parameters $m_i$ one may get some additional constants of motion.
 There are indications of the existence of hidden symmetries in the action-angle variable formulation of the system. For example, if the dependence of the Hamiltonian on two action variables $I_1, I_2$ is of the form $\mathcal{I}=\mathcal{I}(k_1I_1+k_2I_2,\ldots)$, where $k_{1,2}$ are integers (or rational numbers),
 the function $\cos(k_1\Phi_2-k_2\Phi_1)$ (where $\Phi_{1,2}$ are conjugate angle variables) defines a constant of motion additional to the Liouville one, see \cite{GNS-1} and refs therein.
Having the generating function \eqref{Generating-function} at hands, we can get the expressions for action variables and  through them, the expression of the Hamiltonian in terms of elliptic functions.
This analysis, besides its technical difficulty,  is of its own interest and we postpone it to a separate study.

\section{Discussion}

We showed that the particle dynamics on a generic $2N+1$ dimensional NHEMP black hole geometry is integrable by explicitly constructing the generating function \eqref{Generating-function}, extending the results of \cite{GNS-1} for the NHEMP with equal angular momentum parameters (which in our conventions are denoted by $m_i$), to the most general case. Our results establish that the integrability is not a result of the $U(N)$ symmetry of the latter case, which is broken to $U(1)^N$ in the general case. Although in our computations we assumed  non-equal $m_i$ cases, one can show that our results recovers the special cases where some of the $m_i$ are equal. To see the latter, one can study the $m_i-m_j\to 0$ limit for two given $i,j$.

One interesting special case is $m_N=1$. In this case, as \eqref{m-i-a-i} implies $m_i=\infty, \ i\neq N$ and that $r_H=0$. This case hence corresponds to the Extremal Vanishing Horizon (EVH) family \cite{EVH-1, EVH-2} where the near horizon geometry has a (locally) AdS$_3$ part with $SO(2,2)$ isometry. The integrability of course persists in this case too. As another related case one may explore whether the integrability continues over the even dimensional MP black holes.

The techniques we developed in this paper can be used for tackling other problems. Here we mention a few:
\begin{itemize}
\item Although the $SL(2,\mathbb{R})$ isometry appearing in the NH region of extreme black holes was crucially used in our setup, it is plausible that our technical tools are useful in studying causal curves and geodesics around generic (non-extreme) black holes, especially in the near horizon region.
\item One can use the explicit solutions of the Hamilton-Jacobi equations for analyzing Schr\"odinger equation and/or equation of motion of other fields on these backgrounds, before or after taking the NH limit (see \cite{frolov} for a related study). This latter among other things, would be useful for establishing whether the physics of NH is decoupled from the rest of space. Moreover, it is a crucial step toward carrying out quantization of such systems and a systematic study of the quasinormal modes.
\item The information about the NH background geometry and in particular its conserved charges, as we see from our explicit solution \eqref{Generating-function}, is encoded in the geodesics we constructed. On the other hand,  vacuum Einstein equations in higher dimensions allows for solutions with various horizon topologies, e.g. black rings in five or higher dimensions \cite{ring}. It is desirable to explore if the information about horizon topology can also be extracted from our solution.
\item Besides the MP black holes, vacuum Einstein equations admit black hole solutions with $U(1)^{d-3}$ axial isometry. This class of solutions coincide with Kerr and MP black holes respectively in four and five dimensions. The NH geometry of extremal black holes in this class will have $SL(2,\mathbb{R})\times U(1)^{d-3}$ isometry and different aspects of them has been discussed e.g. in \cite{KL-review, NHEGs-1, NHEGs-2}. Based on the experience with generic NHEMP, we expect the conformal mechanics on the NH geometry of these black holes to be integrable. It is interesting to explore this explicitly.
\end{itemize}
Finally it would be interesting to  explore the relevance and significance of our results for the Kerr/CFT proposal \cite{Kerr/CFT} and for the question of hidden symmetries \cite{Hidden}.

\acknowledgments
{
We thank George Pogosyan for useful comments on separation of
variables
in ellipsoidal coordinates.
The work of A.N. and T.H supported in part by Tomsk Polytechnic
University competitiveness enhancement program,
The work of M.M.Sh-J. is supported in part by the SarAmadan of Iran
grant
and also by the junior research chair in black hole physics of Iranian
NSF and he also acknowledges the ICTP Simons
fellowship support and ICTP program NT-04.
}
\appendix
\section{Near-horizon Myers-Perry solution }\label{appendix}

Near-Horizon limit for MP geometries in odd dimensions   is given by the metrics \cite{NHEG-2}
\be\label{metrics}
ds^2= \frac{F_{H}}{b}\left(-r^2d\tau^2+\frac{dr^2}{r^2}\right)+(r^2_H+a^2_i)d\mu_i d\mu_i+\gamma_{ij}D\varphi^i D\varphi^j,,\qquad i,j=1,\cdots , N,
\ee
with the  following notation:
\be
F_H=\sum_{i}\frac{r^2_H}{r^2_H+a^2_i}\mu^2_i,\quad  \gamma_{ij}=(r^2_H+a^2_i)\mu^2_i\delta_{ij}+\frac{1}{F_H}a_ia_j\mu^2_i\mu^2_j,\quad
D\varphi^i \equiv d\varphi^i+k^ird\tau
\ee
where the  constant parameters are defined by the expressions
\be
b=\frac{1}{r^2_H}\left(1+4\sum_{i>j} \frac{r^2_H}{r^2_H+a^2_i}\frac{r^2_H}{r^2_H+a^2_j}\right),
\qquad k_i=\frac{2r_H}{b}\frac{a_i}{(r^2_H+a^2_i)^2} .
\ee
Here  $\mu_i$ are latitude coordinates
\be
\sum_{i=1}^N \mu^2_i=1,
\label{ell0}\ee
$a_i$ are rotational parameters  and $r_H$ is the horizon radius of the black hole  defined by the maximal value of the solution of equation
\be\label{ai}
\sum_i\frac{r^2_H}{r^2_H+a^2_i}=1.
\ee

The above suggests it is convenient to define $m_i$ parameters
\be\label{m-i-a-i}
m_i=\frac{{r^2_H+a^2_i}}{r^2_H}\geq 1, \qquad \sum_{i=1}^N\frac 1m_{i}=1.
\ee
The $N$ independent parameters specifying the system are $m_i, r_H$ (note that $m_i$ provide $N-1$ independent parameters).
We  rescale the latitude coordinates $\mu_i$ introducing the Cartesian coordinate $x_i=\sqrt{m_i}\mu_i$, so that the constraint \eqref{ell0}
the near-horizon metrics \eqref{metrics} takea form \eqref{ell} and \eqref{m2} respectively.

\section{Useful Identities}\label{appendix-useful}
To work through equations in section \ref{sec-conf-mech}, we have used the following identities. Let us consider the set of real numbers $\lambda_a$ where none of them are equal.
Recalling the $(N-1)$th order Lagrange polynomials $l_a(\kappa)$,
\be
l_a(\lambda_b)=\delta_{ab},
\qquad
l_a(\kappa)= \prod_{\substack{
    1 \le b \le N-1 \\
    b\ne a} }
\frac{\kappa-\lambda_b}{\lambda_a-\lambda_b},
\ee
one can prove that for any real constant $\kappa$
\be\label{26}
\frac{1}{\prod_{a=1}^{N-1}(\lambda_a-\kappa)} =   \sum_{a=1}^{N-1}\frac{1}{\prod_{b=1;a\ne b}^{N-1}(\lambda_b-\lambda_a)}\frac{1}{\lambda_a-\kappa}.
\ee
For  $\kappa=0$ we get
\be\label{27}
\frac{1}{\lambda_1\ldots\lambda_{N-1}}= \sum_{a=1}^{N-1} \frac{1}{\prod_{b=1; b\neq a}^{N-1}(\lambda_b-\lambda_a)}\frac{1}{\lambda_a}.
\ee
Noting that the LHS of \eqref{26} is a function with simple roots at $\kappa=\lambda_a$,  \eqref{26} may also be verified using contour integrals over complex $\kappa$-plane. Moreover, one can prove
\be\label{28}
\sum_{i=1}^{N} \frac{\lambda^\beta_i}{\prod_{j=1; i\neq j}^{N}(\lambda_i-\lambda_j)}=\delta_{\beta,
N-1}
\qquad
{\rm for  }\qquad 1\leq  \beta\le N-1.
\ee

\end{document}